\documentclass[]{aastex631}
\usepackage{amsmath}
\usepackage{graphicx}
\usepackage{multirow}
\usepackage{bm}
\usepackage{amssymb}
\usepackage{mathtools}
\usepackage{enumerate}
\usepackage{color}
\usepackage{soul}

\newcommand{\h}{\mathcal{H}}

\begin{document}
	\title{Can the Gravitational Wave Background Feel Wiggles in Spacetime?}
	
	\author{Gen Ye\footnote{ye@lorentz.leidenuniv.nl}}
	\affiliation{Institute Lorentz, Leiden University, P.O. Box 9506, Leiden 2300 RA, The Netherlands}
	 \author{Alessandra Silvestri}
	 \affiliation{Institute Lorentz, Leiden University, P.O. Box 9506, Leiden 2300 RA, The Netherlands}
	
	\correspondingauthor{Gen Ye}
	\email{ye@lorentz.leidenuniv.nl}
	\begin{abstract}
		Recently the international pulsar timing array collaboration has announced the first strong evidence for an isotropic gravitational wave background (GWB). We propose that rapid small oscillations (wiggles) in the Hubble parameter would trigger a resonance with the propagating gravitational waves, leaving unique signatures in the GWB spectrum as sharp resonance peaks/troughs. The proposed signal can appear at all frequency ranges and is common to GWBs with arbitrary origin. The resonant signal can appear as a trough only when the GWB is primordial, and its amplitude will also be larger by one perturbation order than in the non-primordial case. These properties serve as a smoking gun for the primordial origin of the observed GWB. We showcased the viability of the signal to near future observations using the recent NANOGrav 15yr data.
		
	\end{abstract}
%	\maketitle
	\section{Introduction and Conclusion}
	
	A common-spectrum red noise (noise whose power spectrum decreases with increasing frequency) shared between pulsars has been reported in 2020 \cite{NANOGrav:2020bcs,Goncharov:2021oub, Chen:2021rqp}. Recently, multiple pulsar timing array (PTA) collaborations, NANOGrav \cite{NANOGrav:2023gor}, EPTA \cite{Antoniadis:2023ott}, PPTA \cite{Reardon:2023gzh} and CPTA \cite{Xu:2023wog} have announced detection of Hellings-Downs like spatial correlation \cite{Hellings:1983fr} in this common-spectrum signal at $3-4\sigma$, which indicates that such signal is highly likely to be the first detection of an isotropic gravitational wave background (GWB). 
	
	Due to the weakness of the gravitational interaction, GWs, and the GWB as well, are thought to be insensitive to any environmental physics they propagate through, unless such physics has strength beyond the linear regime, which is typically only true near the GW source. As a result, most theoretical studies of the PTA signal has focused on the possible sources such as supermassive black holes, (see~\cite{NANOGrav:2023hfp, Antoniadis:2023zhi} for some up-to-date constraints, and also \cite{Huang:2023chx,Yang:2023aak,Konoplya:2023fmh}), inflation~\cite{Starobinsky:1979ty, Rubakov:1982df,Guzzetti:2016mkm,Vagnozzi:2020gtf, Vagnozzi:2023lwo, Firouzjahi:2023lzg,Unal:2023srk,Niu:2023bsr,Borah:2023sbc,Choudhury:2023kam,Datta:2023vbs}, scalar-induced GWs~\cite{10.1143/PTP.37.831, Matarrese:1992rp, Matarrese:1993zf,Domenech:2021ztg,Franciolini:2023pbf,Cai:2023dls,Wang:2023ost, Yi:2023mbm,Abe:2023yrw,Ebadi:2023xhq,Liu:2023ymk,Zhu:2023faa} and collision of bubbles of first order phase transitions \cite{Kosowsky:1991ua, Caprini:2007xq, Huber:2008hg,Li:2020cjj, NANOGrav:2021flc,Fujikura:2023lkn,Bringmann:2023opz,Addazi:2023jvg,Yang:2023qlf,Zu:2023olm, Megias:2023kiy,Du:2023qvj,Cruz:2023lnq, Xiao:2023dbb, Ashoorioon:2022raz} as well as the leftover topological defects~\cite{Vilenkin:1984ib,Hindmarsh:1994re, Damour:2004kw, Saikawa:2017hiv,Ellis:2023tsl, Gouttenoire:2023ftk,Kitajima:2023cek,Bai:2023cqj,Lazarides:2023ksx,Blasi:2023sej,Wang:2023len, Servant:2023mwt, Li:2023tdx, Lu:2023mcz} and magnetic fields \cite{RoperPol:2022iel,Li:2023yaj}, (see \cite{Caprini:2018mtu} for a review and \cite{NANOGrav:2023hvm, Antoniadis:2023zhi,Bian:2023dnv} for some updated constraints). Very few, on the other hand, consider the propagation effect. Expansion of the Universe on cosmological time scales can influence the broad shape (tilt) of isotropic GWB spectrum, whose effect, however, is highly degenerate with the assumptions on the unknown details of the sources, see e.g.~\cite{NANOGrav:2023hfp, Antoniadis:2023zhi, Liu:2023pau}. Propagation through large scale structure~\cite{Camera:2013xfa,Garoffolo:2019mna, Garoffolo:2022usx} and resonance with local fields~\cite{Degollado:2014vsa,Yoshida:2017cjl,Jung:2020aem,Tsutsui:2023jbk} might also imprint anisotropy (unobserved by now) in the GWB.
	
	Contrary to the previously mentioned belief, in this paper we show that background evolution not only modifies the broad shape of the isotropic GWB spectrum, but also is able to imprint \textit{localized} signature within the reach of near future experiments. The proposed signature is a resonant peak/trough at $k_c=k_{\rm{osc}}/2$ in the GWB spectrum induced by rapid small oscillations (wiggles) with wave number $k_{\rm{osc}}$ of the background spacetime $\h$ on top of its slowly varying part $\langle\h\rangle\equiv\bar{\h}$, i.e.
    \begin{equation}\label{eq:h}
        \h = \langle\h\rangle + \h_{\rm{osc}} \equiv \bar{\h}(1+\delta_{\rm{osc}}),\qquad \delta_{\rm{osc}}<1,
    \end{equation}
    where $\langle\cdot\rangle$ stands for averaging over the rapid oscillation periods. We make no assumptions on the possible physics that sources $\delta_{\rm{osc}}$, except for that the GW is luminal and weakly coupled (which is expected in General Relativity as well as many theories of dark energy/modified gravity). $\delta_{\rm{osc}}$ evades most cosmological observations because only the averaging effect on cosmological time scales is constrained and the effect of background on sub-Hubble process is generally suppressed by powers of $\h/k_{\rm{osc}}$ without resonance. Even for the most precise cosmic probe, i.e. cosmic microwave background (CMB), it is shown that non-negligible oscillatory early dark energy (EDE) is allowed by current observation around recombination \cite{Poulin:2018cxd}. To the best of our knowledge, searching for the proposed resonant signature in the GWB spectrum is the only known method to detect sub-Hubble background spacetime oscillations, possibly sourced by various mechanisms such as early dark energy/modified gravity~\cite{Doran:2006kp,Karwal:2016vyq,Poulin:2018cxd,Zumalacarregui:2020cjh,Braglia:2020iik}, in broad energy scales. In particular, this method allows us to constrain spacetime oscillation from energy scale $T\sim10^9~\rm{GeV}$ (LIGO) down to $T\sim0.1~\rm{eV}$ (CMB B-mode). Because the radiation-dominated Universe is transparent to GWs, it opens up a new window on the dynamics of the Universe long before last-scattering which would otherwise be hard to constrain due to the thermal equilibrium. The resonance mechanism is general and will therefore also influence the scalar sector. Near recombination, it may generate potentially observable contributions in the CMB power spectrum as was shown recently in~\cite{Smith:2023fob}. In this work, we focus on the signatures in the tensor sector.
    
    The resonance mechanism applies to GWBs with arbitrary origin, but the strength and shape of the corresponding signature depend on the latter. It is found that a resonant trough is only possible when the GWB is primordial, and its amplitude will also be larger by one perturbation order than in the non-primordial case. Both properties can serve as a smoking gun for the primordial origin of the GWB. The viability of the signal to near future observations is showcased using the NANOGrav 15yr data.
	
	\section{GW resonance}
	We start from the  propagation equation for a GW mode $\gamma(\tau,k)$ on the homogeneous and isotropic cosmological background
	\begin{equation}\label{eq:gw}
		\ddot{\gamma}+2\Gamma\dot{\gamma}+k^2\gamma=0
	\end{equation}
	where upper dots indicate derivation with respect to the conformal time $\tau$. Here we have promoted the Hubble friction $\h$ to a general friction term $\Gamma$ which can additionally include possible contributions from modified gravity. We assume that the speed of tensors is luminal. In parallel to Eq.\eqref{eq:h}, we can split $\Gamma$ into a part $\bar{\Gamma}$ slowly varying on cosmological time scales plus fast oscillating wiggles $\delta_{\rm{osc}}$ with $k_{\rm{osc}}\gg\h$ 
	\begin{equation}\label{eq:Gamma}
		\Gamma = \bar{\Gamma}(1 + \delta_{\rm{osc}}),\qquad \delta_{\rm{osc}}<1.
	\end{equation}
	% where $\bar{\Gamma}\equiv \langle \Gamma \rangle$ is the average over oscillation periods and generally contains two contributions, $\bar{\h}$ and a possible running of the Planck mass on cosmological scales. 
 As discussed in detail in the Appendix, such wiggles can be sourced in early-dark-energy-like scenarios characterized by a canonical scalar field with a $\phi^4$ potential.
 
  For oscillations of small amplitude, i.e. $\delta_{\rm{osc}}<1$, Eq.\eqref{eq:gw} can be solved perturbatively
	\begin{equation}\label{eq:gw_pt}
		\ddot{\gamma}^{(n)}+2\bar{\Gamma}\dot{\gamma}^{(n)}+k^2\gamma^{(n)}=-2\delta_{\rm{osc}}\bar{\Gamma}\dot{\gamma}^{(n-1)}, \qquad n=0,1,\dots\qquad \gamma^{(-1)}=0.
	\end{equation}
	We will focus on epochs with a constant equation of state $w$ and assume $\bar{\Gamma}\simeq\bar{\h}=\frac{2}{3w+1}\tau^{-1}$; in this case,  the two homogeneous solutions of Eq.\eqref{eq:gw} are:  $\gamma^{(0)}_1\simeq\frac{J_{\nu}(k\tau)}{(k\tau)^{\nu}}, ~\gamma^{(0)}_2\simeq\frac{Y_{\nu}(k\tau)}{(k\tau)^{\nu}},~ \nu=\frac{3(1-w)}{2(1+3w)}$. The leading order correction term $\gamma^{(1)}$, can be found via the  Green's function method
	\begin{equation}\label{eq:gm1_green}
		\begin{aligned}
			\gamma^{(1)}(\tau)&=\int_{\tau_i}^{\tau}-2\delta_{\rm{osc}}\bar{\Gamma}\dot{\gamma}^{(0)}(\tilde{\tau})\frac{\gamma^{(0)}_1(\tilde{\tau})\gamma^{(0)}_2(\tau)-\gamma^{(0)}_2(\tilde{\tau})\gamma^{(0)}_1(\tau)}{\gamma^{(0)}_1(\tilde{\tau})\dot{\gamma}^{(0)}_2(\tilde{\tau})-\dot{\gamma}^{(0)}_1(\tilde{\tau})\gamma^{(0)}_2(\tilde{\tau})}d\tilde{\tau}\\&\simeq\int_{\tau_i}^{\tau}2\frac{\tilde{\tau}^q}{\tau^q}\bar{\Gamma}\delta_{\rm{osc}}\frac{\dot{\gamma}^{(0)}(\tilde{\tau})}{k}\sin[k(\tilde{\tau}-\tau)]d\tilde{\tau}, \quad q=\frac{2}{1+3w}
		\end{aligned}
	\end{equation}
	where in the second line we substituted in the homogeneous solution of $\gamma^{(0)}$ and took the sub-horizon limit by keeping only terms with the highest power in $k\tau$.
	
	The leading order correction, $\gamma^{(1)}$, is generally highly suppressed by both the sub-Hubble parameter $\bar{\h}/k$ as well as rapid oscillations. However it can get amplified if $\delta_{\rm{osc}}$ oscillates in resonance with $\gamma^{(0)}$, in which case its amplitude at resonance will be determined by $|\delta_{\rm{osc}}|$, \textit{regardless of $\bar{\h}/k$}. To see this, let us introduce the ansatz 
    \begin{equation}\label{eq:delta_osc}
        \delta_{\rm{osc}}=\psi_0\sin(k_{\rm{osc}}\tau+\alpha)      
    \end{equation}
    for the wiggles with $\psi_0$ characterizing the amplitude of $\delta_{\rm{osc}}$. The leading order GW with conformal wave number $k$ can be expressed as $\gamma^{(0)}\simeq\gamma_0\sin(k\tau+\beta)/(k\tau)^{q}$. $\alpha$ and $\beta$ are arbitrary phases. Now the  second line of Eq.\eqref{eq:gm1_green} becomes
	\begin{equation}\label{eq:gm1_expr}
		\begin{aligned}
			\gamma^{(1)}(k)&\simeq\frac{4}{1+3w}\frac{\gamma_0\psi_0}{(k\tau)^q}\int_{\tau_i}^{\tau}\frac{1}{\tilde{\tau}}\sin(k_{\rm{osc}}\tilde{\tau}+\alpha)\cos(k\tilde{\tau}+\beta)\sin(k(\tilde{\tau}-\tau))d\tilde{\tau}\\&\simeq\psi_0 f(\Delta k)\frac{\gamma_{0}\cos(k\tau+\alpha-\beta)}{(k\tau)^q}
		\end{aligned}
	\end{equation}
	where we have used the rapid oscillation approximation $\int_{\tau_i}^{\tau}g(\tilde{\tau})\sin(k\tilde{\tau})d\tilde{\tau}\sim\int_{\tau_i}^{\tau}g(\tilde{\tau})\cos(k\tilde{\tau})d\tilde{\tau}\sim\mathcal{O}(1/k(\tau-\tau_i))\to0$ when $k(\tau-\tau_i)\gg 1$ and $\dot{g}/g\ll k$, which is true for sub-Hubble GWs propagating over cosmological distances. Resonance happens at $k_c= k_{\rm{osc}}/2$ and $f(\Delta k)$ characterizes the shape of the resonant peak with $\Delta k\equiv k-k_c$. For $\Delta k/k\ll1$, one has the analytic approximation
	\begin{equation}\label{eq:peakshape}
		f(\Delta k)\sim\frac{1}{1+3w}\left[\rm{Ci}(2|\Delta k|\tau) - \rm{Ci}(2|\Delta k|\tau_i)\right]
	\end{equation}
	where $\rm{Ci}(x)\equiv-\int^\infty_{x}dt\cos(t)/t$ is the cosine integral function. In particular $f(0) = \frac{\log(\tau/\tau_i)}{1+3w}=\frac{\log(a/a_i)}{2}=\frac{N}{2}$, corresponds to one half of the number of e-folds, $N$, the Universe underwent from the beginning of the resonance at $\tau_i$. Specializing to the PTA frequency nHz, if resonance starts at horizon reentry ($k\sim 10^{-9}~\rm{Hz} \sim 10^{5}~\rm{Mpc}^{-1}$) and continues all the way to radiation-matter equality ($k_{\rm{eq}}\sim 10^{-2} ~\rm{Mpc}^{-1}$), we get $N\simeq16$ and $f(0)\simeq8$.
	
	\begin{figure}
		\gridline{\fig{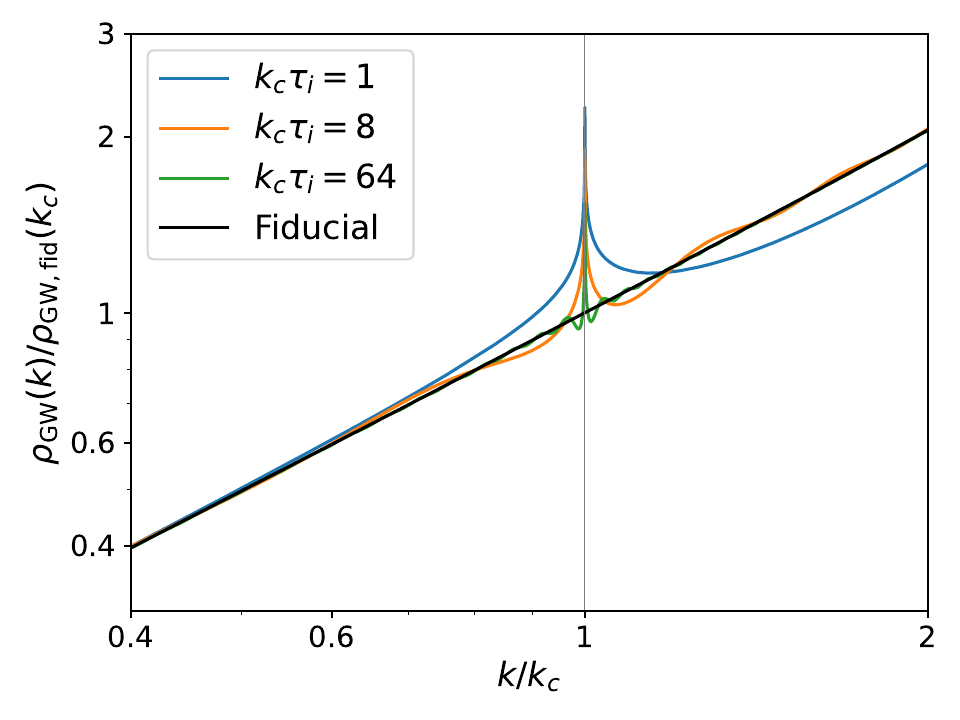}{0.48\textwidth}{(a)}
			\fig{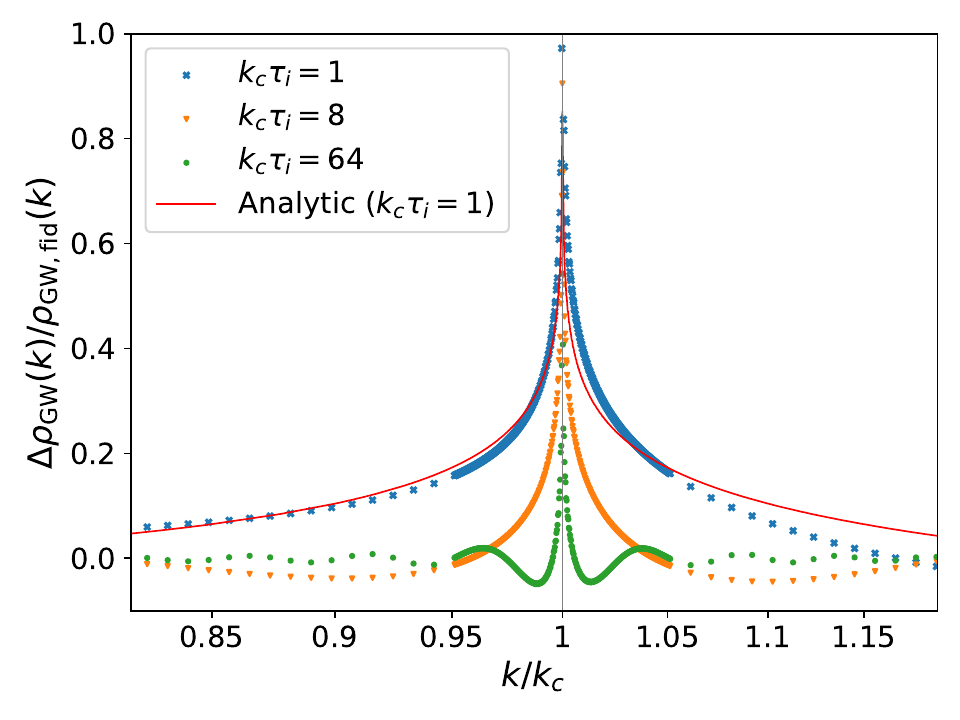}{0.48\textwidth}{(b)}}
		\caption{Full numerical results of energy spectrum of GW with and without (fiducial) resonance for different resonance starting time $k_c\tau_i$. The amplitude of background wiggle is $\psi_0=0.1$. A scale invariant initial GW spectrum and radiation dominance are assumed. In Panel (b) we plot the analytic approximation, from the second line of Eq.\eqref{eq:rhogw} and Eq.\eqref{eq:peakshape}, of the highest peak (i.e. the $k_c\tau_i=1$ case) as a red solid line for comparison.}
		\label{fig:rhogw}
	\end{figure}
	
	\begin{figure}
		\centering
		\includegraphics[width=0.8\linewidth]{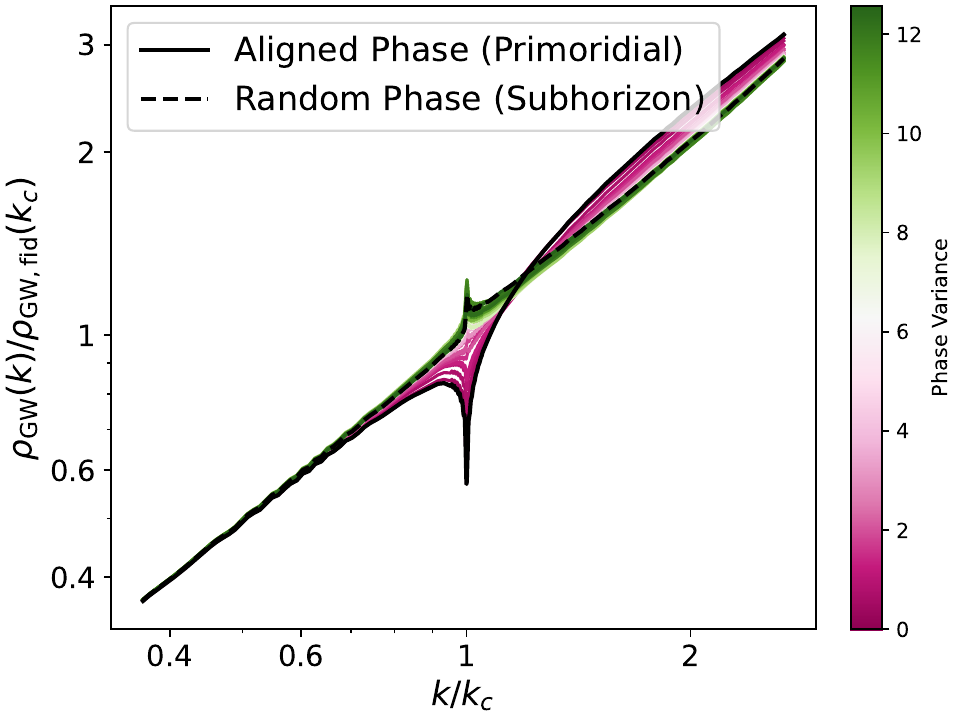}
		\caption{Resonance signal in the GW energy spectrum with a color coding for the variance of the random phase factor. ``Completely aligned phase" corresponds to \textit{Phase Variance} = 0 (reddish color), while ``completely random phase" corresponds to \textit{Phase Variance} $\to\infty$ (greenish color).}
		\label{fig:rhogw_ic}
	\end{figure}
	
	Usually when comparing with observations, the more relevant quantity is the GW energy density spectrum \cite{Boyle:2005se}
	\begin{equation}\label{eq:rhogw}
		\begin{aligned}
			\rho_{GW}(k)&\equiv\frac{d\rho_{GW}}{d\ln k} = \frac{M_p^2}{4}\frac{k^3}{2\pi^2}\left(\left|\frac{d\gamma}{dt}\right|^2+\frac{k^2}{a^2}|\gamma|^2\right)\\&\simeq\left[1+2f(\Delta k)\psi_0\sin(2\beta-\alpha)+f^2(\Delta k)\psi_0^2\right]\frac{M_p^2}{8\pi^2}\frac{k^3}{a^2\tau^2}.
		\end{aligned}
	\end{equation}
	To arrive at the second line, we assumed that only the leading order correction is important, i.e. $\gamma \simeq \gamma^{(0)} + \gamma^{(1)}$ and used the approximation Eq.\eqref{eq:gm1_expr}. The term linear in $\psi_0$ arises from the fact that $\gamma^{(1)}$ is sourced by $\gamma^{(0)}$. There are two physical limits of interest concerning the phase factor $\sin(2\beta-\alpha)$:
	\begin{itemize}
		\item \textbf{Completely random phase} If GWB is of sub-Hubble origin, such as supermassive black hole binaries or bubble/topological defects collision after inflation, the phases $\beta$ of GWs generally satisfy a uniform random distribution. In this case the linear correction term vanishes when summing over all $\beta$ phases, i.e. $\frac{1}{2\pi}\int_0^{2\pi}\rho_{\rm{GW}}d\beta$, and the leading order correction is $\mathcal{O}(\psi_0^2)$. 
		\item \textbf{Completely aligned phase} If GWB undergoes horizon reentry mechanism, such as primordial tensor perturbations generated during inflation, the GWs will have exactly the same phase $\beta\sim\pi/2$ due to the adiabatic initial condition.  $\alpha$ depends on the actual physics that sources the wiggles $\delta_{\rm{osc}}$, thus can take arbitrary values in general. E.g. for the explicit example in the Appendix, $|\alpha|\sim \pi/2$ and $|\sin(2\beta-\alpha)|\sim1$. Therefore we argue here the phase factor in this case does not vanish and that the leading order correction is $\mathcal{O}(\psi_0)$.
	\end{itemize}
	Assuming completely aligned phase, Fig.\ref{fig:rhogw} plots the numeric results of the energy spectrum $\rho_{\rm{GW}}$, defined by the first line of Eq.\eqref{eq:rhogw}. For the case of $k_c\tau_i=1$ we plot also the corresponding analytic approximation from the second line of Eq.\eqref{eq:rhogw}, using Eq.\eqref{eq:peakshape}. One can notice that the analytic approximation is able to capture the overall shape of the peak, with  the peak height slightly smaller than the numerical result. This is to be expected because $\gamma^{(n)}, n\ge2$ are also important at resonance. It is clear from Fig.\ref{fig:rhogw} that the peak width is affected by $k_c\tau_i$. The more sub-Hubble the GW is when resonance starts, the narrower the peak width, though the peak height only depends on how long the resonance lasts logarithmically, through the e-folding number $N$. This implies higher frequency resolution would be required if the resonance starts when the GW is deep sub-Hubble. We will come back to this issue in the next section. It is worth noting here that the signal strength is higher by one perturbation order ($\psi_0$) in the aligned phase case and a resonant trough (negative phase factor) is \textit{only} possible in this case, see the Appendix for a concrete example. Therefore, observation of a resonant trough will immediately imply primordial origin of (part of) the GWB. The difference between a peak and a trough is equivalent to taking $\psi_0\to-\psi_0$, thus without loss of generality, we will always assume a peak hereafter for better presentation. 
	
	A unique property of the resonance signal is that it is sensitive to the origin of the GWB, as explained in the previous paragraph. In particular, a trough is only possible when GWB is primordial. To further illustrate this point, we promote the phase of GWs to a Gaussian random variable with different variance and compute the phase averaged GWB energy spectrum. The results are plotted in Fig.\ref{fig:rhogw_ic}, interpolating between a resonance trough with amplitude $\mathcal{O}(\psi_0)$ to a small peak of order  $\mathcal{O}(\psi_0^2)$. GWB of primordial origin falls into the ``completely aligned phase" case, corresponding to \textit{Phase Variance} = 0. On the other hand, when the GWB originates from many individual and uncorrelated sub-Horizon sources, it falls in the ``completely random phase" case, corresponding to \textit{Phase Variance} = $\infty$ (i.e. uniform distribution).
	
	\section{Resonant GW and PTA}
	\begin{figure}
		
		\gridline{\fig{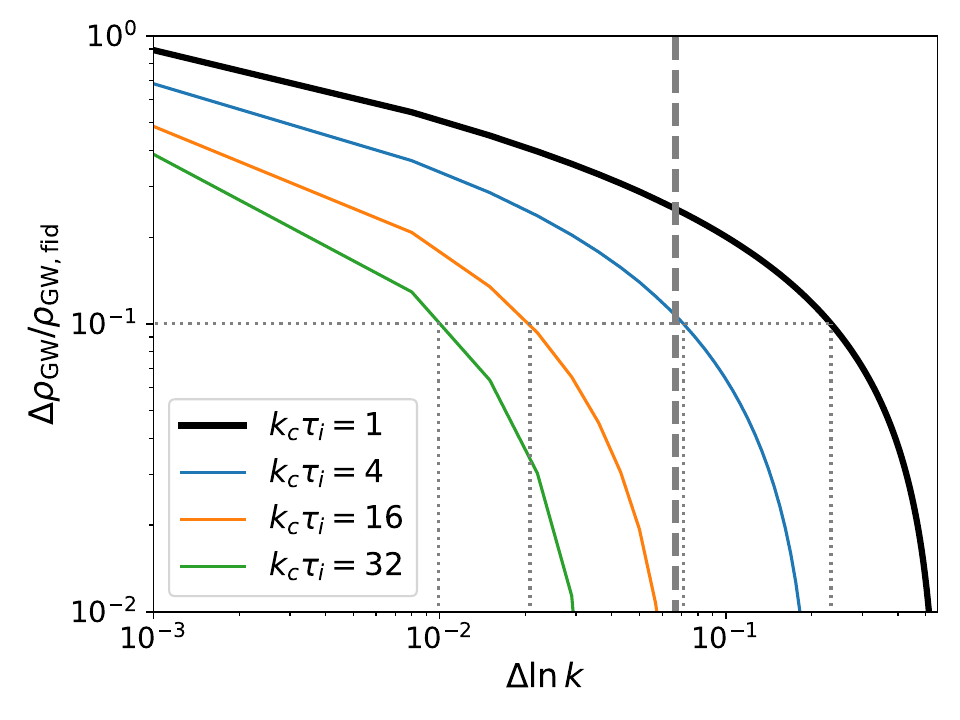}{0.48\linewidth}{(a)}
		\fig{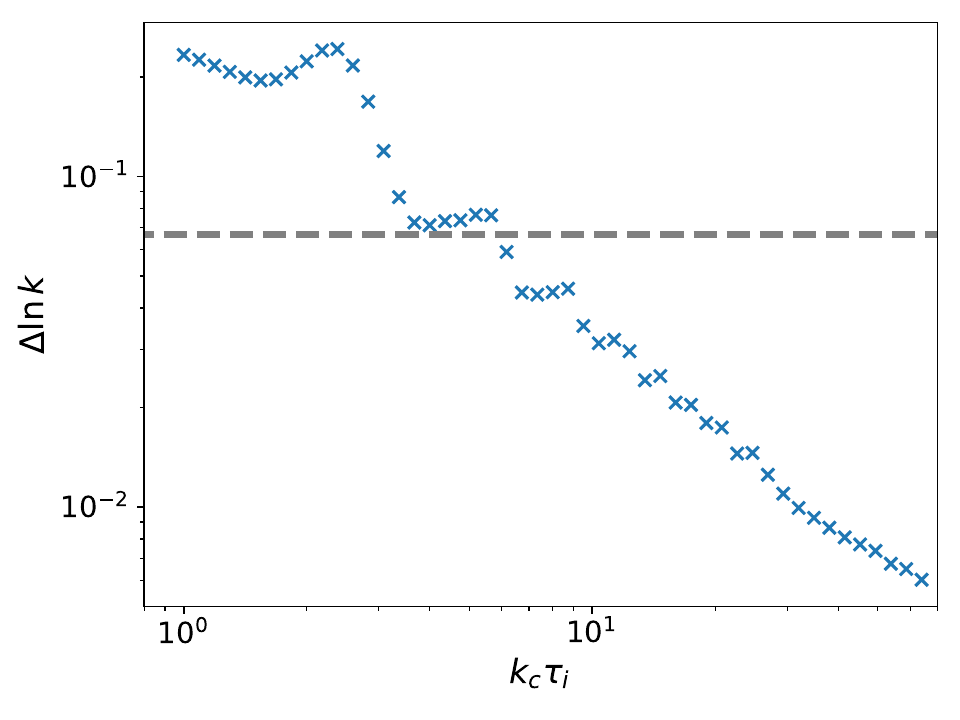}{0.48\linewidth}{(b)}}
		\caption{Integrated power excess signal $\frac{\Delta\rho_{\rm{GW}}}{\rho_{\rm{GW,fid}}}$ induced by GW resonance with background wiggles of amplitude $\psi_0=0.1$, in the logarithmic frequency bin centered at $k_c$. NANOGrav 15yr data detects possible GWB signal up to bin 8 \cite{NANOGrav:2023gor}, corresponding to max frequency resolution $\Delta\ln k\simeq\ln(8/7)/2\simeq0.07$, which is indicated by dashed grey lines in the plots. Black dashed line in the left panel indicates the result for a Gaussian peak with the same peak height and half width as the $k\tau_i=1$ resonant peak (thick black solid line).}
		\label{fig:gw_bandpower}
		
	\end{figure}

 \begin{figure}[h]
	   \centering
			\includegraphics[width=0.8\linewidth]{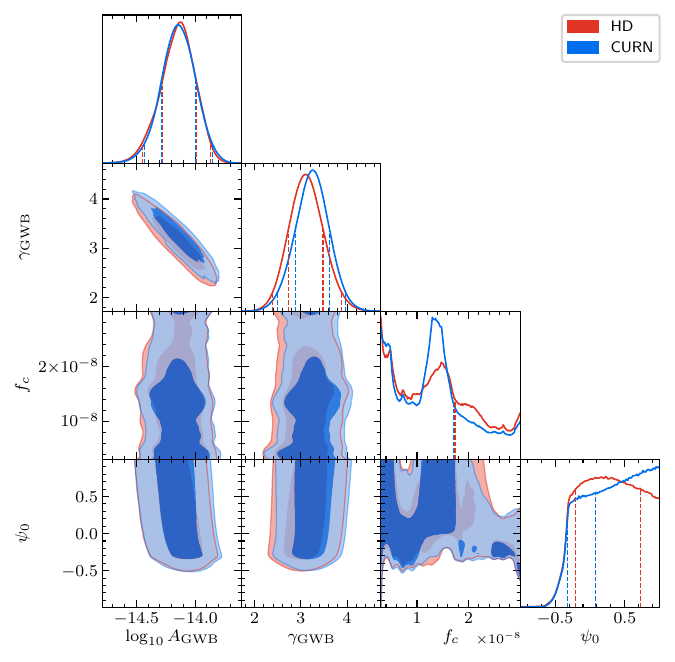}
			\caption{68\% and 95\% posterior distributions of the resonance signal template parameters $\{\log_{10}A_{\rm{GWB}},\gamma_{\rm{GWB}},\psi_0,f_c\}$ fitted to NANOGrav 15yr data.}
		\label{fig:mcmc}
	\end{figure}
	Experiments have finite frequency resolution and measure integrated power in frequency bins. An important property of the integrated signal is that it is insensitive to how long the resonance lasts, i.e. the e-folding number $N$, as long as it is long enough (i.e. $k_c\tau\gg1$ and $\tau/\tau_i\gg1$). This property can be seen by analytically integrating Eq.\eqref{eq:peakshape} over a 
	 frequency bin of width $\lambda\equiv\frac{\Delta k_{\rm{max}}}{k_c}$
  \begin{equation}
      \frac{\int_{|\Delta k|<\lambda k_c}k^3f(\Delta k)d\ln k }{\int_{|\Delta k|<\lambda k_c}k^3d\ln k }\simeq\frac{1}{1+3w}\left[-\rm{Ci}(2\lambda k_c\tau_i)+\frac{\sin(2\lambda k_c\tau_i)}{2\lambda k_c\tau_i}\right] + \mathcal{O}\left(\frac{1}{k_c\tau}\right)
  \end{equation}
	which is independent of $N$. The bin width $\lambda$ should satisfy $(k_c\tau)^{-1}<\lambda\ll1$ for the approximation to hold. The explicit dependence on $k_c\tau_i$ also explains what we see in Fig.\ref{fig:rhogw} regarding the peak width. The resonance peak is very narrow, to assess the viability of Eq.\eqref{eq:rhogw} in observations, we plot the relation between log frequency resolution $\Delta\ln k\equiv\ln(1+\lambda)$ and the signal as integrated power excess $\Delta\rho_{\rm{GW}}/\rho_{\rm{GW,fid}}\equiv(\rho_{\rm{GW}}-\rho_{\rm{GW,fid}})/\rho_{\rm{GW,fid}}$ in the frequency bin $[e^{-\Delta \ln k}k_c, e^{\Delta \ln k}k_c]$ in Fig.\ref{fig:gw_bandpower}. Due to the sharpness of the resonant peak/trough, see Fig.\ref{fig:rhogw}, the integrated power excess has a unique frequency resolution dependence as compared to smooth peaks, i.e. the solid and dashed lines in Fig.\ref{fig:gw_bandpower}, which could be used to distinguish the resonance signal from other possible local signatures. As already mentioned in the previous section, higher frequency resolution is needed to resolve the signal if resonance starts deep sub-Hubble. To see this, we plot in Fig.\ref{fig:gw_bandpower} the required frequency resolution $\Delta \ln k$ to reach $10\%$ power excess signal for different resonance starting time $k_c\tau_i$. According to Fig.\ref{fig:gw_bandpower}, for a detector with 10\% energy precision, the relative frequency resolution needs to be above $40\%$ (note the actual bin size is $2\Delta\ln k$) in order to resolve the resonant signal sourced by wiggles $\delta_{\rm{osc}}$ of order $10\%$. For PTA experiments, if the spectrum is binned with bin width $1/T_{\rm{obs}}$, $T_{\rm{obs}}$ being the observation time span, $40\%$ relative resolution can be reached for bin number $i\ge3$.

\begin{table}[]
    \centering
    \begin{tabular}{c|c}
        Parameter&Prior\\
        \hline
         $\log_{10}A_{\rm{GWB}}$& Uniform[-18, -6]  \\
         $\gamma_{\rm{GW}}$& Uniform[0, 7] \\
         $f_c$ [nHz]& Uniform[3, 30]\\
         $\psi_0$& Uniform[-1, 1]
    \end{tabular}
    \caption{Parameter priors for the MCMC analysis.}
    \label{tab:prior}
\end{table}
 
	Assuming the most optimistic situation, where the resonance starts immediately after horizon reentry, i.e. $k_c\tau_i=1$, and $|\sin(2\beta-\alpha)|=1$, we propose the following resonant GW signal template for PTA
	\begin{equation}\label{eq:gw_template}
		\Omega_{\rm{GW}}(k)=\frac{2\pi^2}{3}\left(\frac{\rm{year}^{-1}}{H_0}\right)^2\left[1+\psi_0F_{\Delta  k}(k, k_c)\right]A_{\rm{GWB}}^2\left(\frac{k/2\pi}{\rm{year^{-1}}}\right)^{5-\gamma_{\rm{GWB}}}
	\end{equation}
	where $\Delta k$ is the width of the frequency bin used and $k_c$ ($f_c$) is the resonance wave number (frequency). Typically $\Delta k\simeq 2\pi/T_{\rm{obs}}$. The shape function $F$ is defined as
	\begin{equation}
		F_{\Delta k}(k, k_c)\equiv\frac{\int_{k-\Delta k/2}^{k+\Delta k/2} \tilde{k}^{1.8}f(\tilde{k}-k)d\ln \tilde{k}}{\int_{k-\Delta k/2}^{k+\Delta k/2} \tilde{k}^{1.8}d\ln \tilde{k}}
	\end{equation}
	where the factor $k^{1.8}$ comes from setting $\gamma_{\rm{GWB}}=3.2$ according to NANOGrav GWB bestfit \cite{NANOGrav:2023gor} so that $\Omega_{\rm{GW}}(k)\propto k^{1.8}$ and $f(\Delta k)$ is approximated by Eq.\eqref{eq:peakshape}. We fit the template Eq.\eqref{eq:gw_template} to the public NANOGrav 15yr data \cite{NANOGrav:2023hde} using Monte Carlo Markov Chain (MCMC) method, with $\Delta k = 2\pi/16.03\rm{yr}$ as described in the paper. The priors are summarized in Tab.\ref{tab:prior}, in which resonance frequency $f_c$ is confined to the region where GWB has been observed \cite{NANOGrav:2023gor}. Fig.\ref{fig:mcmc} shows the MCMC posterior distributions of the spectrum parameters $\{\log_{10}A_{\rm{GWB}}, \gamma_{GWB}, \psi_0, f_c\}$. Current data is not enough to well constrain local features in the spectrum, thus the wide contours in the plot. However, quite interestingly, both the results assuming HD spatial correlation (HD) and common-spectrum uncorrelated red noise (CURN) display a peak in $f_c$ posterior around $15$ nHz, hinting a possible feature there. This peak is significantly higher when CURN is assumed, which leads us to attribute this curiosity to the difference in power excess signal between CURN and HD at frequency bins 6 and 7 (roughly corresponding to $f=10-16$ nHz) in the NANOGrav results \cite{NANOGrav:2023gor}.
	
	\section{Final remarks}
	
	\textit{Yes, it is possible to detect the resonant signal sourced by spacetime wiggles through PTA.} Fitting the template Eq.\eqref{eq:gw_template} to the recent NANOGrav 15yr data hints at a curious feature near $f_c\sim15$ nHz, see Fig.\ref{fig:mcmc}. It is unclear whether such feature is due to physics or statistical fluctuations. Further study is needed to evaluate its credibility.
	
	The proposed resonance signal can appear at all frequencies. The next step would be to study its phenomenology and viability in other frequency bands, corresponding to e.g. ground based LIGO/Virgo/KAGRA \cite{LIGOScientific:2014pky,VIRGO:2014yos,KAGRA:2020cvd}, space based LISA \cite{LISA:2017pwj}, Taiji \cite{Hu:2017mde} and CMB B-mode \cite{Kamionkowski:1996ks,Kamionkowski:1996zd,BICEP:2021xfz}.
	
	\paragraph*{Acknowledgment} \ GY particularly thanks Alice Garoffolo for insightful discussion during the initial phase of this work and for careful proof-read of the draft. PTA data analysis is performed using the \texttt{PTArcade} software \cite{andrea_mitridate_2023_8106173,Lamb:2023jls}. Our work is supported by NWO and the Dutch Ministry of Education, Culture and Science (OCW) (grant VI.Vidi.192.069)
	
	\appendix
	\section{An example model of GW resonance} \label{apdx:phi4}
	
	\begin{figure}[h]
		\gridline{\fig{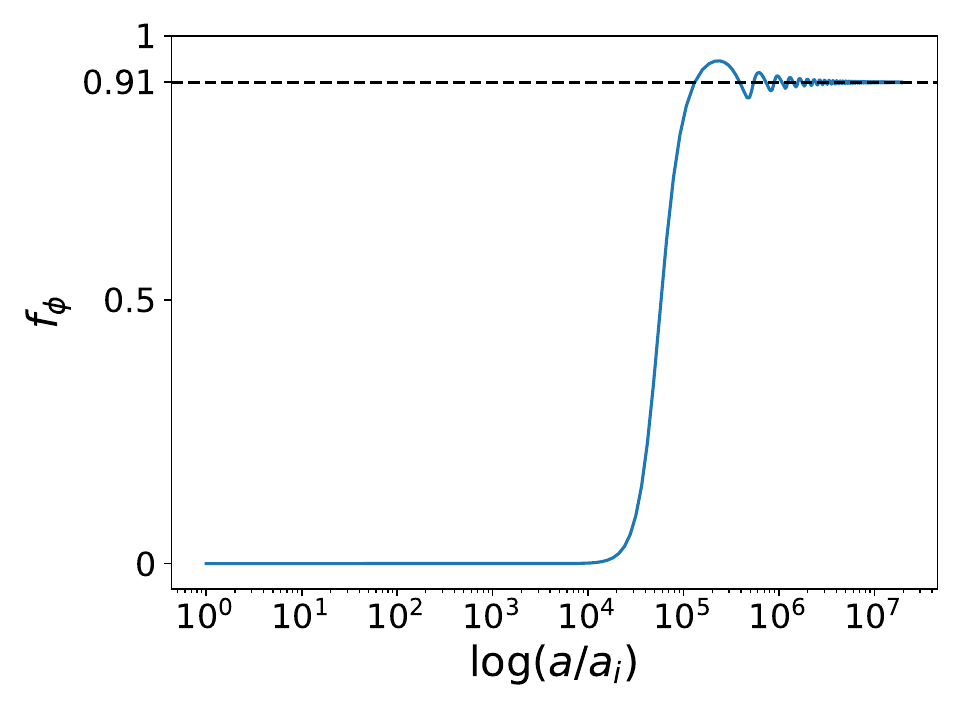}{0.48\textwidth}{(a)}
			\fig{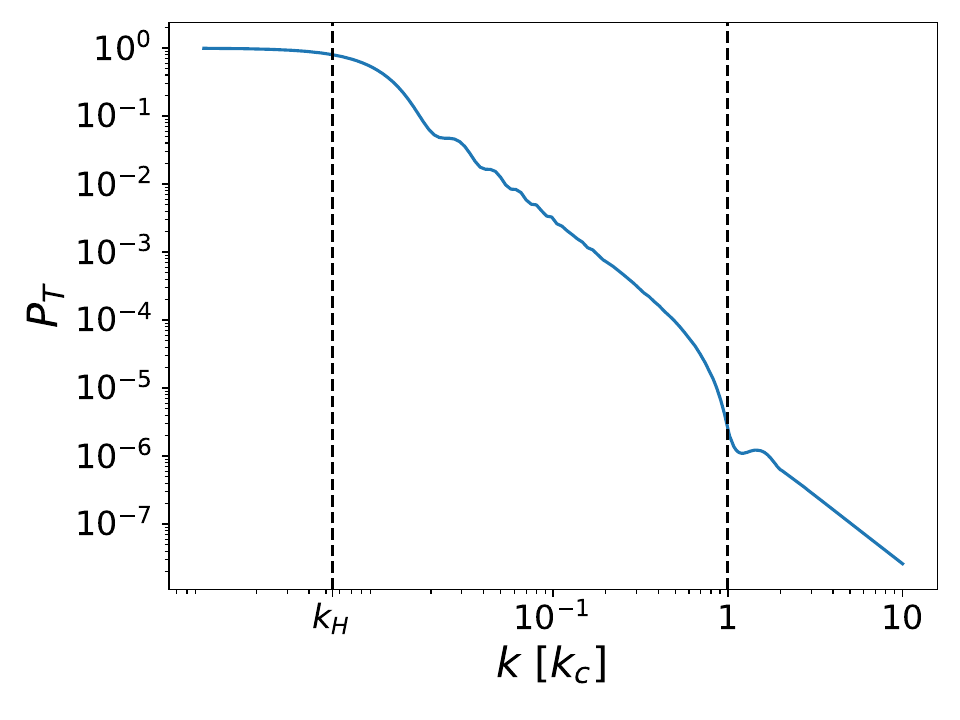}{0.48\textwidth}{(b)}}
		\caption{Numeric results of theory \eqref{eq:action}. (a) Scalar field energy fraction. (b)  Tensor power spectrum. Vertical dashed lines in (b) marks the position of resonance $k_c$ and the conformal Hubble scale $k_H$ at which the spectrum is plotted.}
		\label{fig:phi4}
	\end{figure}
	
	In this appendix we demonstrate that wiggles in $H$ can be sourced by an oscillatory canonical scalar field
	\begin{equation}\label{eq:action}
		\mathcal{S}=\int dx^4\sqrt{-g}\left[\frac{M_p^2}{2}R - \frac{1}{2}(\partial\phi)^2 - \lambda \phi^4\right]+\mathcal{S}_m.
	\end{equation}
	The field is initially frozen at its initial value $\phi_i$ by Hubble friction when $m_{\rm{eff}}^2\sim V_{\phi\phi}\ll H^2$, then thaws at $V_{\phi\phi}\sim H^2$ and undergoes oscillations driven by the potential $V=\lambda\phi^4$ around its minimum. If the thawing time is near matter-radiation equality, such a theory has been shown to be a suitable early dark energy candidate \cite{Agrawal:2019lmo,Ye:2020btb} which significantly alleviates the Hubble tension~\cite{Karwal:2016vyq,Poulin:2018cxd} in a CMB-compatible way, but not fully resolve it~\cite{Hill:2020osr}. The quartic potential  $V(\phi)=\lambda\phi^4$ is important for our example because it drives the radiation-like oscillations in the scalar field. Specifically, at max field displacement $\phi_0$ in one oscillation cycle, one has $V(\phi_0)=\lambda\phi_0^4\sim\rho_{\phi}\propto a^{-4}$ which implies $\phi_0\sim a^{-1}$. At $\phi=0$ we have $\frac{1}{2}\left(\frac{d\phi}{dt}\right)^2\sim\phi_0^2k_{\rm{phys}}^2/2\sim\rho_{\phi}\propto a^{-4}$ implying the physical wave vector $k_{\rm{phys}}\propto a^{-1}$. Thus $\phi$ oscillates with a fixed conformal wave number $k_\phi$, which is essential for resonance with GW. Neglecting the Hubble friction, we can estimate the oscillation period by $T=4\int_0^{\phi_0}d\phi\frac{dt}{d\phi}=4\int_0^{\phi_0}d\phi\left[2\left(V(\phi_0)-V(\phi)\right)\right]^{-1/2}\simeq3.7\lambda^{-1/2}\phi_0^{-1}$. Together with the thawing condition $9H^2\simeq V_{\phi\phi}$ \cite{Marsh:2010wq}, one obtains an order-of-magnitude estimation of the oscillatory frequency
	\begin{equation}\label{eq:k_phi}
		\frac{k_{\phi}}{\mathcal{H}_i}\sim 3 (\lambda f_{\phi})^{1/2}
	\end{equation}
	where we have denoted the conformal Hubble scale at some initial time as $\mathcal{H}_i$ and defined the energy fraction of $\phi$ as $f_{\phi}\equiv\rho_{\phi,i}/3M_p^2H_i^2=\lambda(\phi_{i}/M_p)^4/3$, and in the last equality we de-dimensionalize $\lambda\to \lambda H_i^2M_p^2$. In radiation dominance, $f_\phi=conts.$. Due to time scaling of the scalar field amplitude $\phi\sim a^{-1}$, $\delta_{\rm{osc}}$ is sourced by $\rho_\phi$ through the Freedman equation with $k_{\rm{osc}}=2k_{\phi}$, thus resonance at $k_{c}=k_{\phi}$, with its amplitude proportional to $(\mathcal{H}/k_{\phi})^2$. Therefore, in this case one needs a large $f_{\phi}$ to have a sizable effect, which might appear at the onset of reheating or short scalar dominating phases during radiation dominance. For illustration purpose, we consider $f_{\phi}=0.9$ during radiation dominance and solves the cosmic background and tensor equations numerically. The results are plotted in \ref{fig:phi4}. The plotted modes are initially super horizon with the adiabatic initial condition $h=1, \dot{h}=0$. We observe a resonance trough and because in this example the amplitude of $\delta_{\rm{osc}}$ decays with time, the trough is smoother than that in the main text. For scalar fields, oscillation in $\phi$ can also trigger parametric resonance in the scalar field perturbations, whose phenomenology in the CMB has been studied in Ref.~\cite{Smith:2023fob}.
	
	\bibliography{reference}
\end{document}